\pdfoutput=1
\documentclass[conference]{IEEEtran}

\usepackage{cite}
\usepackage{amsmath,amssymb,amsfonts}
\usepackage{graphicx}
\usepackage{textcomp}
\usepackage{xcolor}
\usepackage{url}
\usepackage{booktabs}
\usepackage{balance}
\usepackage{listings}
\usepackage{tikz}
\usepackage{multirow}
\usepackage{pgfplots}
\usepackage{soul}
\usepackage{eso-pic}
\pgfplotsset{compat=1.18}
\usetikzlibrary{arrows.meta, positioning, shapes.geometric, fit, backgrounds, patterns, calc}

\lstdefinelanguage{Kotlin}{
  keywords={fun, val, var, class, object, if, else, return, const, private, companion, data, sealed, when, in, is, true, false, null, this, it, override, abstract, interface, open, internal, public, protected, import, package, for, while, do, try, catch, finally, throw, as, by, constructor, init, get, set, lazy, lateinit, suspend, coroutine, annotation, typealias, enum},
  morekeywords={String, Int, Boolean, Long, Unit, List, Map, Set, Any, Nothing, Pair, Triple},
  morekeywords=[2]{@Singleton, @Inject, @Named, @DynamicValueOverrides, @VisibleForTesting, @Suppress},
  sensitive=true,
  morecomment=[l]{//},
  morecomment=[s]{/*}{*/},
  morestring=[b]",
}
\lstdefinestyle{kotlinstyle}{
  language=Kotlin,
  basicstyle=\footnotesize\ttfamily,
  keywordstyle=\bfseries,
  commentstyle=\itshape\color{gray},
  stringstyle=\color{gray},
  breaklines=true,
  showstringspaces=false,
  tabsize=2,
  frame=none,
  aboveskip=2pt,
  belowskip=2pt,
  xleftmargin=4pt,
  numbers=none,
}

\begin{document}
\AddToShipoutPictureFG*{%
  \AtPageLowerLeft{%
    \put(\LenToUnit{\dimexpr(\paperwidth-\textwidth)/2\relax},\LenToUnit{0.32in}){%
      \parbox[b]{\textwidth}{\centering\scriptsize
        Accepted for publication in the 2026 IEEE International Conference on Software Maintenance and Evolution (ICSME), Industry Track.\\
        \textcopyright~2026 IEEE. Personal use of this material is permitted. Permission from IEEE must be obtained for all other uses, in any current or future media, including reprinting/republishing this material for advertising or promotional purposes, creating new collective works, for resale or redistribution to servers or lists, or reuse of any copyrighted component of this work in other works.}}}}

\title{Automating Feature Flag Cleanup at Scale\\with Multi-Agent LLM Systems:\\An Industrial Experience at DoorDash}

\author{
\IEEEauthorblockN{Atharv Chandratre}
\IEEEauthorblockA{\textit{DoorDash, Inc.}\\
San Francisco, CA, USA\\
atharv.chandratre@doordash.com}
\and
\IEEEauthorblockN{Jai Datar}
\IEEEauthorblockA{\textit{DoorDash, Inc.}\\
San Francisco, CA, USA\\
jai.datar@doordash.com}
}

\maketitle

\begin{abstract}
  Feature flags are runtime configuration mechanisms used for gradual feature rollouts, experimentation, and
  kill switches. Once a feature is fully rolled out, the flag and its gated code become technical debt and can introduce operational risk through inadvertently enabling dormant code paths. At DoorDash, over 56,000 feature flags span 623 repositories, with 1,659 currently stale as of May 2026, with internal estimates of one to two hours per manual cleanup. Existing rule-based cleanup tools proved insufficient in DoorDash's codebases. Our contribution is an end-to-end automated LLM-based feature flag cleanup system that integrates querying live feature flag rollout data using MCP, Jira ticket processing, and end-to-end pull request creation. Our system successfully produced engineer-approved flag removal PRs for 45 (90\%) of the 50 stale flags it was evaluated on, proving the system's operational viability at scale with each cleanup averaging only \$4.79 and 13.8 minutes. We evaluated the system's quality and reliability boundary based on success rates, cost per cleanup and time metrics. Furthermore, we discuss our lessons learned when architecting this system, to provide practical guidance for developing similar multi-agent systems.

\end{abstract}

\section{Introduction}
\label{sec:introduction}

Feature flags (also known as feature toggles) are a fundamental practice in modern software delivery, enabling safe gradual rollouts, A/B testing, and kill switches~\cite{rahman2016feature, meinicke2020exploratory, hoyos2021removal}. However, once a feature flag has been fully rolled out, the flag becomes \emph{stale}. Its conditional logic no longer serves a purpose and constitutes technical debt. Stale flags also pose a direct operational risk: they preserve dormant code paths that can be inadvertently reactivated. In one well-documented incident, a trading firm repurposed a stale feature flag during deployment, triggering \$460M in unintended trades within 45~minutes~\cite{knightcapital2013}. Despite these risks, prior work has found that approximately 25\% of feature flags persist for over eight years in open-source projects~\cite{meinicke2020exploratory}, and practitioners consistently report that flag cleanup is burdensome and frequently deferred~\cite{hoyos2021removal, mahdavihezaveh2021practices}.

At DoorDash, the problem manifests at a significantly large scale. Our experimentation platform manages over 56,000 Dynamic Values (DVs), the company's feature flag abstraction, across approximately 623 repositories, with 1,659 currently stale. Approximately 2,300 new DVs are created every month. With internal estimates of one to two hours per manual cleanup, the backlog grows faster than teams can address it.

Throughout this paper, we will be using the abbreviation DV to refer to Dynamic Values, the term for feature flags at DoorDash.

The most prominent automated approach, Piranha~\cite{ramanathan2020piranha} and its polyglot successor~\cite{clapp2024polyglotpiranha}, simplifies conditional expressions by matching AST patterns. As we detail in Section~\ref{sec:background:flags} and quantify in Section~\ref{sec:background:motivation}, DoorDash's dependency-injected wrapper pattern, non-boolean DVs, and cross-file cleanup requirements make these tools structurally inapplicable in our setting.

To address these challenges, we designed and deployed a multi-agent LLM system built on Google ADK~\cite{google2025adk} that automates the full lifecycle of feature flag cleanup at DoorDash. An orchestrator agent analyzes each stale flag, queries the experimentation platform for its live production rollout state via MCP (Model Context Protocol)~\cite{anthropic2024mcp}, and formulates a cleanup strategy. It then spawns specialized removal agents that execute code changes in parallel, each operating in an isolated \emph{git worktree}~\cite{git2024worktree} to prevent interference between concurrent cleanups. The pipeline autonomously handles Jira ticket processing, code modification, test validation, and pull request creation, producing merge-ready PRs requiring zero manual intervention between the initial engineer confirming the inferred target value and final PR creation.

Our contributions are:
\begin{itemize}
\item The design and deployment of a multi-agent architecture at scale for feature flag cleanup, where an orchestrator coordinates parallel removal agents operating in isolated git worktrees.
\item An end-to-end automated pipeline performed by each cleanup agent that autonomously bridges issue tracking, querying the experimentation platform using MCP, code modification, test validation, and pull request creation for every stale feature flag.
\item A detailed industrial evaluation of our LLM-based feature flag cleanup system, reporting success rates, cost, time, complexity breakdown, and failure analysis, along with an analysis of why rule-based tools such as Piranha do not apply to DoorDash's DV access pattern (Section~\ref{sec:background:flags}). This evidence is absent from literature we reviewed~\cite{nikolov2025google, duolingo2026agents, relevanceai2026byebyeflag, reflag2025cleanup, glovo2025flags, atlassian2026rovo}.
\end{itemize}

\section{Background \& Motivation}
\label{sec:background}

\subsection{Feature Flags at DoorDash }
\label{sec:background:flags}

DVs follow a lifecycle: creation $\rightarrow$ experimentation $\rightarrow$ rollout $\rightarrow$ stale $\rightarrow$ removal. A DV is classified as stale when it: (1) has not been modified for 90 days (2) is still referenced in code (3) is not in an end-of-lifecycle status (archived or retired) (4) is not on an exclusion list. Jira tickets are automatically created for stale DVs identified by a daily cron job and assigned to the DV's creator.

DVs implementation follows a three-layer access pattern in Doordash services, as illustrated in Listing~\ref{lst:dv-pattern}. Layer~1 defines the DV name as a string constant in a Runtime keys file. Layer~2 provides a dependency-injected wrapper function in a feature flag file that calls the \texttt{DynamicValuesClient} API (\texttt{getBoolean}, \texttt{getString}, or \texttt{getObject}) with context dimensions such as user ID or country code. Layer~3 is business logic that injects the \texttt{FeatureFlags} singleton via dependency injection and calls the wrapper.

\begin{lstlisting}[style=kotlinstyle, caption={Example three-layer DV access pattern (simplified).}, label=lst:dv-pattern, float=t]
// Layer 1: RuntimeKeys.kt - string constant
object RuntimeKeys {
  const val ENABLE_X = "enable_feature_x"
}
// Layer 2: FeatureFlags.kt - wrapper function
@Singleton
class FeatureFlags @Inject constructor(
    private val dvClient: DynamicValuesClient
) {
  fun shouldEnableX(userId: String): Boolean =
    dvClient.getBoolean(ENABLE_X,
      withContext("user_id" to userId), false)
}
// Layer 3: Business logic - usage via DI
class MyService @Inject constructor(
    private val featureFlags: FeatureFlags
) {
  fun process(userId: String) {
    if (featureFlags.shouldEnableX(userId)) {
      handleNewPath()
    } else { handleOldPath() }
  }
}
\end{lstlisting}

This pattern, while providing clean separation of concerns, means that the DV name, the API call, and the conditional logic exist in \emph{different files} connected only through method calls and dependency injection. These are semantic relationships that AST-based tools cannot follow.

\subsection{Why Automation Is Needed}
\label{sec:background:motivation}

With 1,659 stale DVs and internal estimates of one to two hours per manual cleanup, clearing the backlog would require thousands of engineering hours. Approximately 2,300 new DVs are created monthly, ensuring the problem compounds faster than teams can address it.

A critical challenge is \emph{target value determination}. When a DV has been fully rolled out, the code default may differ from the production value. A tool operating on source code alone could silently hardcode the wrong value. Our agent queries the experimentation platform via MCP to retrieve the live rollout state, ensuring the correct value is used when performing the cleanup.

Cleaning up a DV requires removing the constant definition, the wrapper function, all call sites (replacing with the target value or simplifying conditional logic), updating or removing affected test cases, and ensuring the CI/CD pipeline passes. Even for a simple boolean DV, this involves coordinated changes across 3-20 files. These changes require understanding code semantics, not just syntax.

\section{System Design}
\label{sec:system}

\subsection{Two-Phase Architecture}
\label{sec:system:architecture}

\begin{figure*}[t]
\centering
\includegraphics[width=0.9\textwidth, height=0.3\textheight]{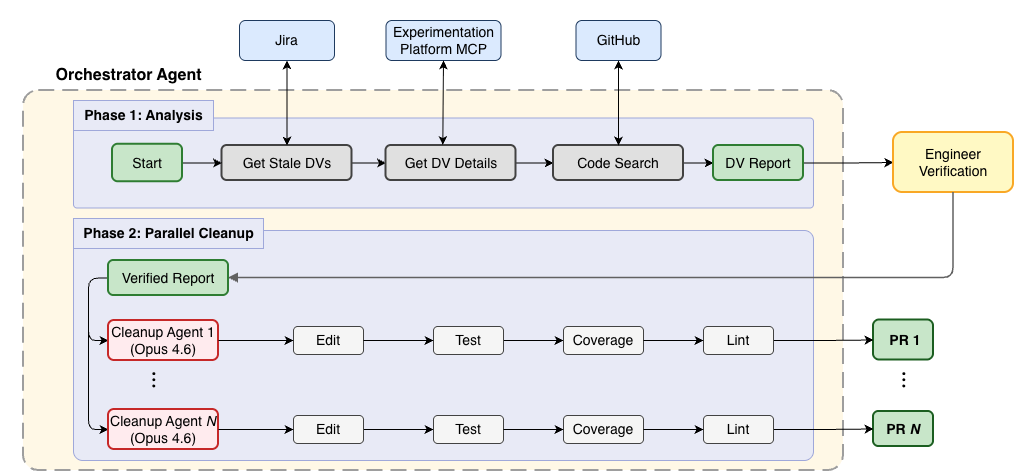}
\caption{System architecture. The orchestrator agent (dashed rectangle) manages the cleanup lifecycle in two phases. In Phase~1, it retrieves stale DVs from Jira, queries the experimentation platform for rollout state, and searches the codebase for all references. The resulting DV report is reviewed by an engineer who confirms target values. In Phase~2, the orchestrator spawns parallel cleanup agents in isolated git worktrees, each editing code, running tests, verifying coverage, and linting before creating a pull request.}
\label{fig:architecture}
\end{figure*}

Figure~\ref{fig:architecture} shows the system architecture. The system operates in two phases with a human-in-the-loop checkpoint.

\textbf{Phase~1 (Analysis \& Report).} The orchestrator agent, powered by Claude Sonnet~4.6~\cite{anthropic2025claude}, processes stale DV cleanup requests. For each DV, it: (1)~fetches the STALEDV Jira tickets via CLI, (2)~queries the experimentation platform via MCP for DV metadata (UUID, rollout percentage, target value), (3)~discovers the local repository and searches for all code references, and (4)~generates a structured confirmation report listing the proposed target value and affected files for each DV.

The experimentation MCP integration is noteworthy: by calling \texttt{GetDynamicValue} and \texttt{GetDynamic\-Value\-Details} endpoints, the agent retrieves the current production rollout state, including rollout percentage and winning variant. This enables correct target value determination. For example, a DV introduced for a feature that was ultimately abandoned will reflect a 0\% rollout state in the platform, ensuring the agent removes it using the baseline value rather than assuming a rolled out flag.

\textbf{Human Checkpoint.} The engineer verifies each DV's proposed target value before cleanup proceeds. This acts as a safety check and ensures DVs with ambiguous rollout states, such as partial rollouts or abandoned experiments are correctly handled.

\textbf{Phase~2 (Parallel Cleanup).} For each confirmed DV, the orchestrator spawns a dedicated, isolated removal agent powered by Claude Opus~4.6~\cite{anthropic2025claude} in a git worktree~\cite{git2024worktree}. Up to four agents run concurrently per repository (configurable value). Each agent executes the iterative cleanup workflow described in Section~\ref{sec:system:removal}, resulting in a pull request assigned to the DV's owner.

\subsection{Removal Agent Design}
\label{sec:system:removal}

Each removal agent follows an iterative workflow: (1)~search for all DV references in the codebase, including the constant definition, wrapper function, and all call sites; (2)~determine a cleanup strategy based on DV type (boolean, integer, string) and usage pattern; (3)~apply edits, replacing wrapper calls with the target value and simplifying conditional logic; (4)~remove the wrapper function, definition and constant; (5)~update or remove all affected test cases (e.g., deleting ``DV disabled'' test branches, removing \texttt{@DynamicValueOverrides} annotations); (6)~ensure build passes (7) verify tests pass (8) verify patch coverage $\geq$95\%; (9)~run the Detekt linter; and (10)~if validation fails, diagnose and fix the issue before retrying.

\textbf{Safety mechanisms.} Four mechanisms prevent incorrect changes: (a)~\emph{git worktree isolation}: each agent operates in a dedicated worktree, preventing interference between concurrent agents; (b)~a \emph{one-hour hard timeout} per agent, preventing runaway sessions; (c)~\emph{JaCoCo patch coverage} validation ensuring $\geq$95\% coverage on changed lines; and (d)~\emph{Detekt static analysis} enforcing code style and detecting common issues. Gradle is executed with \texttt{--no-daemon} to eliminate cross-worktree state pollution during parallel operations.

\section{Evaluation}
\label{sec:evaluation}

\subsection{Research Questions and Setup}
\label{sec:eval:setup}

We evaluate our system along three dimensions:
\begin{itemize}
\item \textbf{RQ1 (Deployment Outcome):} How often does the deployed system produce PRs that engineers can merge directly or after minor revision?
\item \textbf{RQ2 (Reliability Boundary):} What feature flag characteristics predict whether the system produces a merge-ready cleanup in a single shot, requires minor revision, or needs engineer intervention?
\item \textbf{RQ3 (Operational Viability):} What are the time and cost characteristics of running the system in practice? Can it run at scale?
\end{itemize}

\textbf{Dataset.} We analyzed the 50 most recent stale DV removals performed by our system (as of 4/14/26) across multiple Kotlin-based repos at DoorDash. We classified these 50 DVs into three complexity levels: \emph{simple} (single boolean check, 1-2 files; $n$=6), \emph{medium} (multiple usages, 3-5 files, some test changes; $n$=18), and \emph{complex} (nested conditionals, cross-file dependencies, non-boolean types, or significant test updates; $n$=26). This classification was done post hoc using these criteria. DV types include boolean~(41), string~(6), and integer/long~(3).

\textbf{Methodology.} For each DV, we ran the agent and evaluated the generated pull request on three criteria: CI passes (including 95\% patch coverage), code correctness, and developer review. We classify outcomes as: \emph{Accepted} (PR accepted in the first shot without changes); \emph{Accepted with revision} (accepted after minor fixes, typically one additional agent iteration); or \emph{Failed} (engineer had to step in for significant corrections).

\subsection{Results}
\label{sec:eval:results}

\textbf{RQ1 (Deployment Outcome).} Table~\ref{tab:results} summarizes the evaluation results. Of the 50~DVs, 31 (62\%) were cleaned up successfully in the first shot, 14 (28\%) required minor revisions, and 5 (10\%) failed. The overall cleanup rate, defined as DVs where the agent produced a usable PR, was 90\%.

Of the 14~revisions, the most common cause was insufficient patch coverage (6~cases), where the agent needed one additional iteration to add or update tests. The remaining 8 involved incomplete dead code removal, typically missing a dead variable or reference in a deep call chain, which the agent fixed when reprompted.

CI passed on the first attempt for 44 of 50~DVs (88\%). The 6~initial failures were all coverage-related; none were due to incorrect code changes. This aligns with our system design: the agent is constrained from creating a PR unless unit tests and static analysis (Detekt) pass locally. While the agent's system prompt explicitly mandates 95\% patch coverage locally, we observed occasional miscalculations in its LLM-driven self-assessment. DoorDash's CI pipeline caught these discrepancies as it fails if patch coverage is $<$95\%.

\begin{table}[t]
\centering
\caption{Evaluation results on 50 stale DVs.}
\label{tab:results}
\begin{tabular}{lr}
\toprule
\textbf{Metric} & \textbf{Value} \\
\midrule
DV cleanups analyzed & 50 \\
Accepted (first-shot) & 31 (62\%) \\
Accepted with revision & 14 (28\%) \\
Failed & 5 (10\%) \\
\midrule
Overall cleanup rate & 45/50 (90\%) \\
CI first-pass rate & 44/50 (88\%) \\
\midrule
Avg.\ time per DV & 13.8 min \\
Median time per DV & 12.3 min \\
Avg.\ LLM API cost per DV & \$4.79 \\
Median LLM API cost per DV & \$3.49 \\
\bottomrule
\end{tabular}
\end{table}

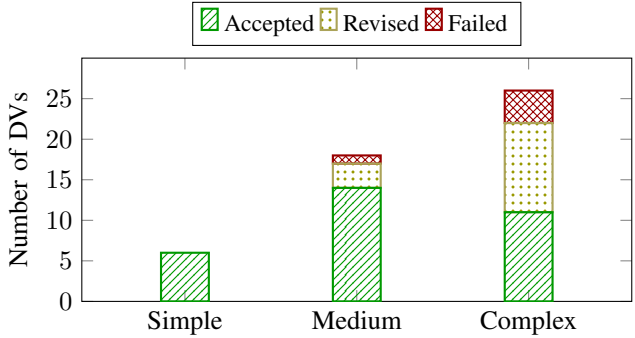
\begin{figure}[t]
\centering
\begin{tikzpicture}
\begin{axis}[
    ybar stacked,
    bar width=18pt,
    width=\columnwidth,
    height=4.8cm,
    ylabel={Number of DVs},
    symbolic x coords={Simple, Medium, Complex},
    xtick=data,
    legend style={at={(0.5,1.05)}, anchor=south, legend columns=3, font=\small},
    ymin=0, ymax=30,
    ytick={0,5,10,15,20,25},
    enlarge x limits=0.3,
    every axis plot/.append style={thick},
]
\addplot+[fill=green!35, pattern=north east lines, pattern color=green!60!black, draw=green!60!black] coordinates {(Simple,6) (Medium,14) (Complex,11)};
\addplot+[fill=yellow!40, pattern=dots, pattern color=yellow!60!black, draw=yellow!60!black] coordinates {(Simple,0) (Medium,3) (Complex,11)};
\addplot+[fill=red!25, pattern=crosshatch, pattern color=red!60!black, draw=red!60!black] coordinates {(Simple,0) (Medium,1) (Complex,4)};
\legend{Accepted, Revised, Failed}
\end{axis}
\end{tikzpicture}
\caption{Cleanup outcomes by complexity level. Simple DVs achieve 100\% first-shot acceptance; complex DVs show higher revision and failure rates but still achieve 85\% overall cleanup.}
\label{fig:outcome-bar}
\end{figure}

Figure~\ref{fig:outcome-bar} breaks down results by complexity. All simple DVs were cleaned up in a single pass. Medium-complexity DVs achieved a 94\% cleanup rate (17/18). Even for complex DVs involving cascading parameter removal through 5+ method layers, nested conditionals, and significant test refactoring, the overall cleanup rate was 85\% (22/26).

\textbf{RQ2 (Reliability Boundary).} Figure~\ref{fig:outcome-bar} breaks down outcomes by complexity. All simple DVs
  (n=6) produced merge-ready PRs in a single pass. Medium-complexity DVs achieved a 94\% cleanup rate: 14 of 18
   in a single pass, 3 after one revision (2 for incorrect test modifications, 1 for insufficient patch
  coverage), and 1 requiring engineer intervention due to incomplete removal. Complex DVs present the
  reliability boundary at an 85\% cleanup rate: 11 of 26 succeeded in a single pass, 11 required one revision
  (4 for incorrect test modifications, 2 for residual unused variables, 5 for insufficient patch coverage),
  and 4 required engineer intervention due to incomplete removal. This is because removing Complex DVs is characterized by cascading
  parameter removal through 5+ method layers, nested conditionals, and significant test refactoring. The
  distinguishing factor across all 5 intervention cases is call-chain depth with cross-interface parameter
  threading, not flag type or file count. In each case, the agent correctly identified and removed the
  majority of dead code but missed one or two links in a multi-file chain; the remaining work was completable
  by an engineer. Critically, across all 50 DV removal PRs generated, we did not observe any bugs or regressions introduced by the changed code, indicating that the system's reliability boundary is one of completeness,
  not correctness.

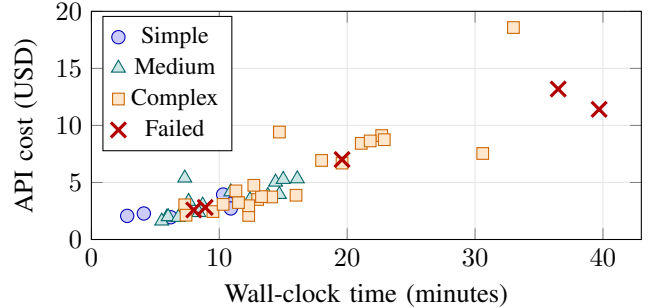
\begin{figure}[t]
\centering
\begin{tikzpicture}
\begin{axis}[
    width=\columnwidth,
    height=4.6cm,
    xlabel={Wall-clock time (minutes)},
    ylabel={API cost (USD)},
    xmin=0, xmax=43,
    ymin=0, ymax=20,
    legend style={at={(0.02,0.98)}, anchor=north west, font=\small},
    grid=major,
    grid style={gray!20},
]
\addplot[only marks, mark=*, mark size=2.5pt, blue!70!black, fill=blue!20]
  coordinates {(10.7,3.17)(10.9,2.72)(6.2,1.95)(2.8,2.07)(10.3,3.94)(4.1,2.28)};
\addplot[only marks, mark=triangle*, mark size=3pt, teal!80!black, fill=teal!20]
  coordinates {(8.7,3.02)(13.8,3.85)(14.4,5.03)
  (5.9,1.96)(16.1,5.32)(7.3,5.39)(5.5,1.62)(12.9,3.50)
  (14.7,3.95)(7.6,3.34)(8.4,2.36)(12.4,3.49)(6.0,2.00)
  (6.8,1.94)(11.6,3.30)(15.0,5.27)(10.9,4.18)};
\addplot[only marks, mark=square*, mark size=2.2pt, orange!80!black, fill=orange!20]
  coordinates {(16.0,3.88)(9.5,2.45)(14.1,3.73)(14.7,9.42)
  (12.3,2.11)(13.0,3.51)(12.3,2.96)(11.5,3.24)(19.6,6.70)
  (7.3,3.04)(21.1,8.43)
  (22.7,9.11)(13.3,3.76)(30.6,7.54)(11.3,4.26)(21.8,8.64)
  (10.3,3.08)(7.4,2.13)(33.0,18.58)(18.0,6.93)(12.7,4.75)
  (22.9,8.75)};
\addplot[only marks, mark=x, mark size=4pt, very thick, red!70!black]
  coordinates {(39.7,11.41)(8.9,2.81)(8.0,2.59)(36.5,13.19)(19.6,7.01)};
\legend{Simple, Medium, Complex, Failed}
\end{axis}
\end{tikzpicture}
\caption{Agent cost vs.\ wall-clock time for each DV, colored by complexity. Failed DVs are marked with $\times$. Simple and medium DVs cluster in the low-cost region; complex DVs show higher variance.}
\label{fig:scatter}
\end{figure}

\textbf{RQ3 (Operational Viability).} Figure~\ref{fig:scatter} plots wall-clock time against API cost for all 50~DVs, colored by complexity. Simple and medium DVs cluster in the low-cost, low-time region, while complex DVs exhibit wider variance with outliers reaching 40~minutes and \$19. The 5~failed DVs tend toward longer execution times. Table~\ref{tab:cost-by-complexity} breaks down averages by complexity.

\begin{table}[t]
\centering
\caption{Time and cost by DV complexity.}
\label{tab:cost-by-complexity}
\begin{tabular}{lrrrrr}
\toprule
\textbf{Complexity} & \textbf{$n$} & \textbf{Avg.\ time} & \textbf{Med.\ time} & \textbf{Avg.\ cost} &
\textbf{Med.\ cost} \\
\midrule
Simple & 6 & 7.5 min & 8.2 min & \$2.69 & \$2.50 \\
Medium & 18 & 10.4 min & 9.9 min & \$3.46 & \$3.42 \\
Complex & 26 & 17.7 min & 14.4 min & \$6.20 & \$4.50 \\
\bottomrule
\end{tabular}
\end{table}

The agent processed DVs in an average of 13.8~minutes (median: 12.3~min) at \$4.79 (median: \$3.49) per cleanup. Internal engineering estimates put manual cleanup at one to two hours of focused effort per DV. Even at the conservative end of this range, the agent's runtime represents a substantial reduction in engineering effort. This demonstrates that the system has the time and cost characteristics to run at scale and saves a substantial amount of time and money when compared to an engineer performing feature flag cleanup manually.

\section{Discussion \& Lessons Learned}
\label{sec:discussion}

\textbf{Model role specialization yields cost-effective results.} We use Claude Sonnet~4.6~\cite{anthropic2025claude} for orchestration (requiring broad context management) and Claude Opus~4.6~\cite{anthropic2025claude} for code modification (requiring deep reasoning about code semantics). This split balances cost and capability: orchestration involves lightweight metadata gathering and planning, while code modification requires deep reasoning about semantics, call chains, and test dependencies. By reserving the more expensive model for the cognitively demanding editing phase, we reduce overall API cost without sacrificing cleanup quality.

\textbf{Git worktree isolation is essential for parallel agents.} Each removal agent operates in an isolated worktree, preventing interference between concurrent cleanups. A failed cleanup is safely discarded by removing the worktree. Running Gradle with \texttt{--no-daemon} prevents daemon conflicts across worktrees. Early experiments without worktree isolation led to race conditions where concurrent agents modified the same files, producing corrupted diffs.

\textbf{Experimentation platform integration prevents silent regressions.} Without access to live rollout state, any automated cleanup approach, whether rule-based or LLM-based, risks hardcoding incorrect values and preserving the wrong behavior after flag removal. In our setting, this integration added negligible latency while eliminating a recurring source of target-value errors, making it one of the highest-leverage design choices in the system.

\textbf{Developer feedback.} We collected informal feedback from engineers who reviewed agent-generated PRs. One engineer noted: \emph{``The agent got every dead code line and dead unit test removed. Given the size of this PR, that's quite impressive.''} Another observed a code quality limitation: \emph{``The removal was correct, but variable names should have been updated, since the flag's removal changed their semantics.''} A third described a common pattern across revised PRs: \emph{``Everything worked in the first commit apart from patch coverage. After one reprompt the agent added the missing tests and it was good to merge.''}

For teams considering similar automation, our results suggest three practical prerequisites: a reliable source of deployed flag state, worktree isolation for parallel edits, and deterministic validation gates before PR creation.

\textbf{Threats to validity.} \emph{Internal validity:} Each DV was processed once; LLM non-determinism means re-running could yield different outcomes. Complexity labels were assigned post hoc by the authors based on the classification described in Section~\ref{sec:eval:setup}. The one to two hour manual cleanup estimate per feature flag is based on internal engineering experience rather than precise wall-clock measurement. \emph{External validity:} Our evaluation covers repositories at a single company with a specific DV access pattern (wrapper-function access via dependency injection). Organizations using direct flag API calls where tools like Piranha are more directly applicable may see different results. The analysis was also performed on feature flags removed in Kotlin-based repos at DoorDash. Generalizability to other languages, flag frameworks, or codebases requires further study. 

\section{Related Work}
\label{sec:related}

\textbf{Feature flag management.} Prior work shows that feature flags accumulate as technical debt and that cleanup practices are inconsistently followed in practice~\cite{rahman2016feature, meinicke2020exploratory, hoyos2021removal, mahdavihezaveh2021practices, rahman2023chromium}. These studies establish both the persistence of stale flags and the diversity of flag-usage patterns in large codebases.

\textbf{Automated code cleanup.} Piranha~\cite{ramanathan2020piranha} is the most directly comparable prior system, automatically generating cleanup diffs for 1,381~stale flags at Uber. Its polyglot successor~\cite{clapp2024polyglotpiranha} extends support to multiple languages via a unified rule engine. These tools target settings where explicit AST rewrite rules can simplify flag-gated code; our setting differs due to dependency-injected wrapper pattern, non-boolean DVs, and cross-file refactoring including test maintenance. Other automated cleanup tools address adjacent but distinct problems: dead code removal (SCARF~\cite{shackleton2023scarf}), large-scale AST refactoring (ClangMR~\cite{wright2013clangmr}, OpenRewrite~\cite{moderne2024openrewrite}), LLM-based dead code elimination at snippet level (DCE-LLM~\cite{chen2025dcellm}), and LLM-validated technical debt remediation (ACE~\cite{tornhill2025ace}). None target end-to-end feature flag cleanup.

\textbf{LLM-assisted flag cleanup in industry.} Several organizations have recently deployed LLM-based systems for feature flag removal, including Google (Gemini)~\cite{nikolov2025google}, Duolingo (Codex~CLI)~\cite{duolingo2026agents}, Relevance~AI (Claude)~\cite{relevanceai2026byebyeflag}, Reflag (GPT-4o)~\cite{reflag2025cleanup}, Glovo (Copilot)~\cite{glovo2025flags}, and Atlassian (Rovo~Dev)~\cite{atlassian2026rovo}. These reports describe system architectures and qualitative outcomes but do not publish quantitative evaluation with success rates, cost, time, complexity breakdown, failure analysis, or comparison against rule-based baselines. The closest architectural analog is bye-bye-flag~\cite{relevanceai2026byebyeflag}, which shares our use of parallel agents in git worktrees and rollout-based target value derivation; our system differs in model role specialization, deterministic patch coverage enforcement, and MCP-based experimentation platform integration. Our work also provides a rigorous empirical evaluation of this class of systems.

\section{Conclusion}
\label{sec:conclusion}

We presented an LLM-based system for automated feature flag cleanup at DoorDash that combines orchestration, parallel removal agents, experimentation-platform lookup, and workflow-integrated validation. On the 50 most recent stale Dynamic Values processed by the system, it produced usable pull requests for 45 flags at an average cost of \$4.79 and 13.8~minutes per flag. Our evaluation covers success rates, cost, time, complexity breakdown, failure modes, and a comparison against rule-based baselines, providing empirical evidence absent from prior industry reports of LLM-based flag cleanup. Our lessons learned highlight that model role specialization, repository
  isolation, live platform integration, and deterministic validation gates are
  practical prerequisites for deploying LLM-based code cleanup at scale. Future work includes several directions: (1)~introducing a confidence score in Phase~1 to auto-approve low-risk cleanups - e.g., boolean DVs at 100\% rollout where the code default matches the production value. This reserves the human checkpoint only for ambiguous cases; (2)~adding a post-cleanup code quality review step to catch semantic issues such as variable names that no longer reflect their purpose after flag removal; (3)~extending support to additional languages; and (4)~integrating with CI pipelines to trigger cleanup automatically at staleness thresholds.

\newpage
\balance

\bibliographystyle{IEEEtran}
\bibliography{references}

\end{document}